\documentclass[conference,a4paper]{IEEEtran} 
\usepackage{graphicx} 
\usepackage{cite}     
\usepackage[hidelinks]{hyperref}
\usepackage{amsmath}
\usepackage{tabularx}
\usepackage{booktabs}
\usepackage{multirow}
\usepackage{xcolor}
\usepackage{makecell}
\usepackage{caption}
\usepackage{enumerate}
\usepackage[acronym,nonumberlist]{glossaries}

\newcommand{\Autoref}[1]{Section~\ref{#1}}

\makeglossaries

\newacronym{ml}{ML}{Machine Learning}
\newacronym{bdt}{BDT}{Binary Decision Tree}
\newacronym{dt}{DT}{Decision Tree}
\newacronym{rf}{RF}{Random Forest}
\newacronym{arf}{ARF}{Adaptive Random Forest}
\newacronym{vfdt}{VFDT}{Very Fast Decision Tree}
\newacronym{hls}{HLS}{High Level Synthesis}
\newacronym{fpga}{FPGA}{Field Programmable Gate Array}
\newacronym{gpu}{GPU}{Graphics Processing Unit}
\newacronym{dse}{DSE}{Design Space Exploration}
\newacronym{hdl}{HDL}{Hardware Description Language}
\newacronym{fxp}{FxP}{FiXed-Point}
\newacronym{fp}{FP}{Floating-Point}
\newacronym{dos}{DoS}{Denial of Service}
\newacronym{adwin}{ADWIN}{ADaptive WINdowing}
\newacronym{asic}{ASIC}{Application Specific Integrated Circuit}
\newacronym{fgt}{FGT}{ForeGround Tree}
\newacronym{bgt}{BGT}{BackGround Tree}
\newacronym{rtl}{RTL}{Register Transfer Level}
\newacronym{axis}{AXIS}{Advanced eXtensible Interface Streaming}
\newacronym{axi}{AXI}{Advanced eXtensible Interface}
\newacronym{axil}{AXIL}{Advanced eXtensible Interface Lite}
\newacronym{bram}{BRAM}{Block RAM}
\newacronym{uram}{URAM}{Ultra RAM}
\newacronym{lfsr}{LFSR}{Linear Frequency Shift Register}
\newacronym{dsp}{DSP}{Digital Signal Processing}
\newacronym{lut}{LUT}{Look-Up Table}
\newacronym{ff}{FF}{Flip Flop}
\newacronym{soc}{SoC}{System on Chip}
\newacronym{dma}{DMA}{Direct Memory Access}
\newacronym{ps}{PS}{Processing System}
\newacronym{hpb}{HPB}{High Performance Bus}
\newacronym{uart}{UART}{Universal Asynchronous Rx/Tx}
\newacronym{cpu}{CPU}{Central Processing Unit}
\newacronym{sram}{SRAM}{Static Random-Access Memory}
\newacronym{lsram}{LSRAM}{Large SRAM}
\newacronym{usram}{uSRAM}{Micro SRAM}
\newacronym{auc}{AUC}{Area Under the Curve}
\newacronym{hbm}{HBM2}{High Bandwidth Memory 2}
\newacronym{tmr}{TMR}{Triple Modular Redundancy}
\newacronym{ddr}{DDR}{Double Data Rate}
\newacronym{mac}{MAC}{Multiply ACcummulate}
\newacronym{ghcordic}{GH-CORDIC}{Generalized Hyperbolic Coordinate Rotation Digital Computer}
\newacronym{dp}{DP}{Dot Product}
\newacronym{sfxp}{SFxP}{Signed FiXed-Point}
\newacronym{ie}{IE}{Inference Element}
\newacronym{wagi}{WAGI}{Weighted Average Gini Impurity}
\newacronym{gp}{GP}{Grace Period}
\newacronym{pe}{PE}{Processing Element}
\newacronym{cnn}{CNN}{Convolutional Neural Network}

\begin{document}

\title{OpenWoud: Vendor-Unlocked Scalable Adaptive Random Forest Accelerator}

\author{Ewout van Nimwegen\IEEEauthorrefmark{1}, Elijah Seth Cishugi\IEEEauthorrefmark{1}, Hendrik Folmer\IEEEauthorrefmark{1}, Marco Ottavi\IEEEauthorrefmark{1}\IEEEauthorrefmark{2} and Kuan-Hsun Chen\IEEEauthorrefmark{3},
\\
\IEEEauthorrefmark{1}University of Twente, The Netherlands,  
\IEEEauthorrefmark{2}University of Rome Tor Vergata, Italy, 
\IEEEauthorrefmark{3}University of Paderborn, Germany\\
\IEEEauthorrefmark{1}\{e.w.vannimwegen, e.s.cishugi, h.h.folmer, m.ottavi\}@utwente.nl, \IEEEauthorrefmark{3}kuan-hsun.chen@uni-paderborn.de
\\\vspace{-20pt}
}

\maketitle
\pagestyle{empty}


\begin{abstract}
Online \gls{ml} enables real-time decision-making on evolving data streams. These dynamic models learn continuously to adapt to new concepts, as previous ones become invalid, but their performance is often limited by throughput. The \gls{arf} is a state-of-the-art model for this setting, which naturally exposes parallelism and is therefore well-suited for hardware acceleration. However, prior \gls{fpga}-based accelerators are vendor-locked and rely on resource-hungry \gls{fp} arithmetics, limiting their scalability and portability. We present \emph{OpenWoud}, a vendor-unlocked \gls{arf} framework for \glspl{fpga}. Our design replaces \gls{fp} operators with a flexible \gls{fxp} arithmetic library and introduces fully transparent, manually scheduled dataflows for training and drift detection. This enables fine-grained time/area trade-offs and predictable behavior across platforms. A comprehensive evaluation on both synthetic and real datasets shows that OpenWoud maintains or increases accuracy (up to +12.89\%), compared to a prior \gls{fpga} accelerator, while reducing the resource utilization by half. Competitive performance is observed compared to GPU (0.5 - 1.5$\times$ speed-up) and an \gls{fpga} baseline (0.2 - 0.5$\times$), despite the latter being based on idealized simulation. Crucially, OpenWoud sustains true sample-by-sample inference, unlike GPUs that require batch insertion.
\end{abstract}


\section{Introduction}

%
%
Satellite jamming detection and network intrusion detection are 
applications that require immediate action on events 
that cause loss of communication \cite{boumeftah_adaptive_2025} \cite{salim_cybersecurity_2025}. 
These are examples of evolving data streams, in which 
the environment changes partly or completely, invalidating 
the previously learned model.
This phenomenon is known as concept drift.
Traditional batch-based offline \gls{ml} models, such as \glspl{cnn} \cite{li_survey_2022} 
or transformers \cite{agarwal_transformers_2023}, 
are designed for static datasets and must re-train from scratch upon changes in the data.
This introduces long delays that a real-time system cannot afford.

Online stream-based \gls{ml} offers a dynamic alternative.
The model learns continuously, by processing a single sample at a time,
for which it returns a prediction and 
trains on the sample, to provide an updated model for the next prediction.
A popular state-of-the-art algorithm for this classification task is the \gls{arf}
\cite{gomes_adaptive_2017} \cite{zhang_farf_2021} 
\cite{alqabbany_measuring_2021}.
It consists of an ensemble of fast decision trees that learn
independently and intuitively require a parallel architecture to 
increase throughput. This makes the \gls{arf} well-suited to port to 
a hardware accelerator \cite{wu_gpu-based_2020} \cite{ridder_accelerated_2023}. However, existing FPGA-based accelerators face two major limitations. First, they are vendor-locked to Xilinx FPGAs, preventing cross-vendor deployment. This is limiting because the design cannot exploit alternative fabric technologies—for example, vendors such as Microsemi offer flash-based FPGAs (as opposed to SRAM-based technology from Xilinx) that provide relatively low-cost, radiation-hardened devices for improved fault-resilience in aerospace applications (e.g., satellites). Second, they rely on \gls{fp}-based arithmetics in 
dataflows, which consumes significant \gls{fpga} resources, limiting
the scalability.

To address this, we present \emph{OpenWoud}, a vendor-unlocked
scalable ARF framework for FPGAs.
OpenWoud replaces complex \gls{fp}-based arithmetics 
(division, square root, natural logarithm) in the train and drift-detector 
dataflows with flexible \gls{fxp} alternatives, enabling fine-grained time/area trade-offs.
%
%
This design is scalable on two fronts:  locally, through parameterized arithmetic operations; and globally, through hyperparameter tuning (e.g., tree count, depth and class count). 
%
%
Our quantitative evaluation (benchmark test) shows that OpenWoud
maintains or increases accuracy (up to 12.89\%), 
achieves competitive throughput, and reduces the area footprint by half.
The contribution of this work is threefold:

\begin{itemize}
    \item We present a vendor-unlocked ARF
        accelerator, improving its usability without hindering its precision.
    \item A flexible open-source FxP arithmetic library,
    with fine-grained control over precision and resource utilization.
    \item We propose a scalable training and drift-detection dataflow architecture,
        which is fully transparent and uses manually derived, adjustable task schedules.
\end{itemize}

The remainder of this paper is organized as follows:
\Autoref{sec:background} introduces the \gls{arf} and \gls{fpga} acceleration,
while \Autoref{sec:relatedwork} covers existing accelerators.
\Autoref{sec:design} describes the OpenWoud framework,
its flexible \gls{fxp} arithmetic library and scalable dataflows. 
Then \Autoref{sec:implementation} details the implementation and
hyperparameter alignment across models. 
\Autoref{sec:evaluation} provides a comprehensive evaluation.
Finally, \Autoref{sec:conclusion} concludes the paper.
\section{Background}\label{sec:background}

This section introduces the \gls{arf} and supporting concepts 
to efficiently map it onto an \gls{fpga} hardware accelerator.

\subsection{Adaptive Random Forest}

The \gls{arf} \cite{gomes_adaptive_2017} is a state-of-the-art algorithm to learn 
efficiently from evolving data streams and handle concept drift. 
The algorithm consists of an ensemble of \glspl{vfdt} and a majority voter.
VFDTs split nodes using the Hoeffding bound to determine \emph{when} (in time) and the
information gain to decide on \emph{what} attribute, to learn efficiently from infinite streams 
\cite{hulten_mining_2001}. However, a single \gls{vfdt} is also prone to instability and easily biased. The \gls{arf} follows the \gls{rf} \cite{breiman_random_2001} to address this issue by combining the knowledge of the diverse ensemble of weak learners
(trains each tree on a random subset of attributes), masking biased trees and causing the overall error to decrease. Specifically, two additional mechanisms are used to extend the \gls{rf} for evolving data streams:

(1) Online bagging \cite{oza_online_2001}: introduces randomness between trees (diversity) through effective re-sampling 
using weights generated with a Poisson distribution.

(2) Drift detector: pairs each tree with a concept drift detection mechanism,
to sense and recover from environment changes.
On a change, the poor performing tree is replaced with another, trained only on the new concept.

\subsection{Concept drift detection}

\begin{figure}[t]
    \centering
    \includegraphics[width=0.925\linewidth]{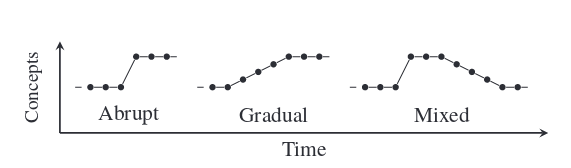}
    \caption{Types of concept drift \cite{ridder_accelerated_2023}.}
    \label{fig:typesOfDrift}
\end{figure}

In \glspl{arf}, concept drift detection is typically implemented using the
\gls{adwin} \cite{bifet_learning_2007}
\cite{grulich_scalable_2018} algorithm, which captures three drift types 
(see \autoref{fig:typesOfDrift}), namely, \emph{abrupt}, \emph{gradual}, and \emph{mixed} drift. \emph{Abrupt} drift occurs at once,
while \emph{abrupt} drift occurs more slowly (spread over tens of thousands of samples, e.g., 25k - 50k).
\emph{Mixed} drift contains both sudden and gradual patterns.
The error rate of each tree is stored in its own dynamic sliding window.
A statistical cut-check is periodically applied by splitting the sliding window into two sub-windows
and testing for significant differences in their data distributions. 
When a change is detected, the older portion is discarded, so 
the detector adapts to the new concept.

\subsection{FPGA acceleration}

Trees in an ARF ensemble operate independently, yielding intrinsic parallelism. FPGAs are a strong fit for ARF hardware acceleration because they support true sample-by-sample processing, exploit fine-grained parallelism, and efficiently handle streaming data, whereas GPUs are typically limited to batch-based sample insertion.
However, the training and drift detection components
contain algorithms with high design complexity:

\begin{itemize}
\item Hoeffding bound (train):
$
\epsilon = \sqrt{\frac{R^2 \cdot \ln(1/\delta)}{2\cdot n}}
$ 

\item Information gain (train):
$
G_{WA} = \sum_{b}^{B} \frac{T_b}{T} \cdot (1 - \sum_{c=1}^{C} P_c^2)
$ 

\item Cutting point (drift):
$
\epsilon_{cut} = \sqrt{\frac{2}{m} \cdot \sigma_W^2 \cdot \ln \frac{2}{\delta'}} + \frac{2}{3m} \ln \frac{2}{\delta'}
$
\end{itemize}
FPGA fabrics do not include hardware resources for complex arithmetic operations like division, square root, and natural logarithm, but only for addition and multiplication. Naive implementations of these operations often consume excessive resources or incur high latency. Hence, both the arithmetic primitives and the overall algorithm architecture must be designed carefully.

\subsection{State-of-the-Arts}\label{sec:relatedwork}

Prior work includes both GPU and FPGA accelerators 
based on the original, CPU-targeted, \gls{arf} model by Gomes et al \cite{gomes_adaptive_2017}; These constitute the baselines against which we evaluate our work.

%
%
\paragraph{ARF (GPU baseline)~\cite{wu_gpu-based_2020}}
It accelerates an ARF using a hybrid GPU–CPU architecture, \emph{GSARF}, reporting a $138\times$ speedup over their CPU baseline; however, this improvement applies only to batch-based sample insertion.
Inference complexity is reduced using dataset preprocessing,
replacing attribute values with a boolean. Moreover, the throughput is increased by approximating the information gain. \emph{Candidate trees} are introduced to reduce training effort, during concept drift recovery on recurring concepts.

%
%
\paragraph{ARF [FPGA baseline]~\cite{ridder_accelerated_2023}}
As the first reported FPGA-based ARF, the design inherits the two \emph{GSARF} optimizations and outperforms the GPU approach by $1.6$–$7\times$ (evaluated in simulation), all while processing the stream sample-by-sample. However, the framework is vendor-locked to Xilinx FPGAs. It was designed using Vitis-HLS tools to target the Alveo U55. The arithmetic operations within the training dataflow and the drift-detection control path were reliazed using proprietary floating-point libraries, the latter being the dominant performance bottleneck in ARF. The resulting FP architectures have compiler-generated task schedules with limited transparency and only coarse control via pragmas, often ignored or inconsistently applied by Vitis~\cite{folmer_high-level_2025,folmer_high-level_2025-1}.
\section{Architecture Design}\label{sec:design}

This section discusses the design of 
\emph{OpenWoud}\footnote{https://anonymous.4open.science/r/OpenWoud-866E}, 
an ARF hardware accelerator for FPGAs, 
designed to overcome the vendor-dependency (restricted deployment to Xilinx-FPGAs) 
and limited scalability (due to \gls{fp}-based arithmetic and compiler-generated,
opaque dataflows) of 
the FPGA baseline by Ridder et al \cite{ridder_accelerated_2023}.
Our design introduces (i) a vendor-unlocked ARF framework,
(ii) a flexible \gls{fxp} arithmetic library, and (iii) 
scalable dataflows for training and drift detection, with manually derived, fully transparent task schedules.

\subsection{High-level overview}

\begin{figure}[t]
    \centering
    \includegraphics[width=\linewidth]{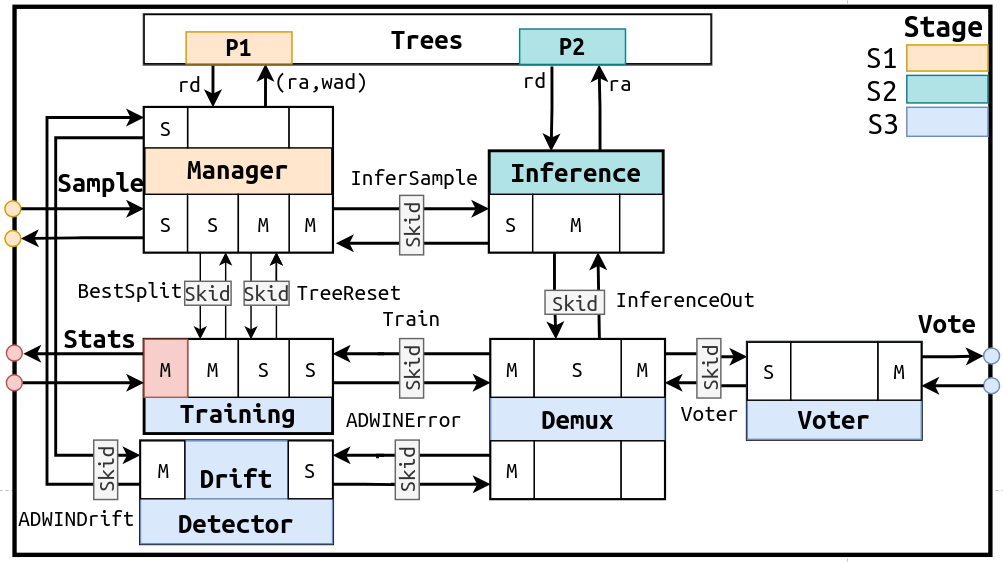}
    \caption{OpenWoud high-level overview. 
    Protocol definitions are \gls{axis} slave: S, master: M, \gls{axi} highlighted in red.}
    \label{fig:arf_system_design}
\end{figure}

OpenWoud processes the evolving data stream in three stages: (S1) management, (S2) inference and
(S3) training/drift detection/voting.
The high-level components (\autoref{fig:arf_system_design}) are connected through 
\gls{axis} interfaces, for a two-way communication flow, 
allowing stages to run concurrently.

%
%
\textbf{S1:} The \textbf{manager} centrally controls the forest by distributing incoming samples across all trees,
inserting splits,
resetting trees when concept drift is detected,
and tracking each tree's state (active/inactive) and role (foreground/background).

%
%
\textbf{S2:} Next, the \textbf{inference-engine} traverses each tree from root to leaf,
using a pipelined architecture to enable parallel processing of the forest.
Through the inference-demultiplexer, leaf end-points are passed on to the training-engine, 
while predictions are forwarded to the majority voter,
and the error-rate goes to the drift detector.

%
%
\textbf{S3:} The final stage consists of 
training/drift detection/voting operating in parallel:
\begin{itemize}
    \item \textbf{Training-engine:} computes splits from leaf
    statistics (\emph{branch counters, branch class counters and total counter}), 
    with FxP-based dataflow architecture to process the  
    information gain and hoeffding bound. 
    Splits are reported back to the manager.
    \item \textbf{Drift detector:} implements ADWIN2 \cite{bifet_learning_2007} 
    to sense concept drift independently for each tree.
    Using a FxP-based dataflow for the cutting points,
    applied on the histogram-based sliding window, for efficient error tracking.
    Drift warnings/errors are reported back to the manager.
    \item \textbf{Voter:} tracks votes from foreground trees and 
    sends the majority class to the output stream.
\end{itemize}

Tree nodes are stored in local dual-port \gls{bram}, while leaf statistics 
go to main memory using multiple \gls{axi} interfaces.
Skid buffers decouple the critical path introduced by backpressure, improving
timing. 
Unlike the FPGA baseline, relying on Vitis-HLS, 
OpenWoud is implemented in Clash an open-source \gls{hdl}
\cite{baaij_clash_2009}.
In contrast to \gls{hls} that insert timing and autogenerate task schedules, 
\glspl{hdl} require explicit timing definition
\cite{lahti_high-level_2025}.
The Clash-compiler generates generic Verilog, ensuring cross-FPGA-platform compatibility, to enable deployment on Lattice and Microsemi targets.

\subsection{Flexible arithmetic library}

Efficiently mapping complex arithmetic operations (division, square root, natural logarithm) 
is key to construct 
scalable train and drift detection dataflows.
The FPGA baseline~\cite{ridder_accelerated_2023}
relies on static FP-based arithmetic operations from Vitis-HLS,
with limited flexibility.
To address this, we developed an open-source \gls{fxp} library based on existing algorithms,
using Clash. 
Its key features: (i) dynamic (yet scalable) resolution, 
(ii) multiple configurations optimized for time/area.

We conducted a thorough \gls{dse}:
8 division, 11 square root and 7 natural logarithm existing FPGA-optimized FxP-based 
implementations are considered.
Our \gls{dse} narrowed down to \emph{iterative fast} methods, these
both scale and achieve high performance, as the required resolutions
throughout the data paths are unknown (not relevant for FP in the FPGA baseline).
Our dataflows use the Newton-Raphson \cite{libessart_scaling-less_2017} 
\cite{pandey_fixed-point_2021}
method for division and square root,
a multiplication intensive algorithm with fast convergence.
While the natural logarithm is implemented using \gls{ghcordic} 
\cite{bangqiang_base-n_2013} \cite{llamocca_fixed-point_2006}, 
only requiring shift/add operations.
The library provides two configurable modes: \emph{multi-cycle} is area-optimized
and \emph{pipelined} is performance-optimized, enabling time/area trade-offs.
To maximize efficiency, we applied FPGA optimizations:

\begin{itemize}
    \item \textbf{Efficient normalization:} single cycle leading-one-detector 
    \cite{libessart_scaling-less_2017} for normalization, reducing the iterations.
    \item \textbf{Hardware efficient data types}: normalized operants use Q2.15 to 
    match hardware specific $18\times 25$ and $18\times 18$ \gls{dsp} units from Xilinx and Microsemi respectively. 
    Reducing hardware multiplier usage by $4\times$ compared to FP-based implementations.
    \item \textbf{Iteration tuning:} square root/division/natural logarithm converge in 4/6/8 iterations 
    respectively for 32-bit output precision, ensuring increased accuracy over
    the FPGA baseline \cite{ridder_accelerated_2023}.
\end{itemize}

\subsection{Dataflows}

Both training and drift detection dataflows are explicitly 
defined to ensure full transparency and flexibility,
rather than compiler-generated schedules.
We reformulate the information gain, Hoeffding bound, and cutting point 
into FPGA-optimized variants, e.g., removing constant terms
and moving divisions outside summations.

\begin{figure}[t]
    \centering
\includegraphics[width=\linewidth]{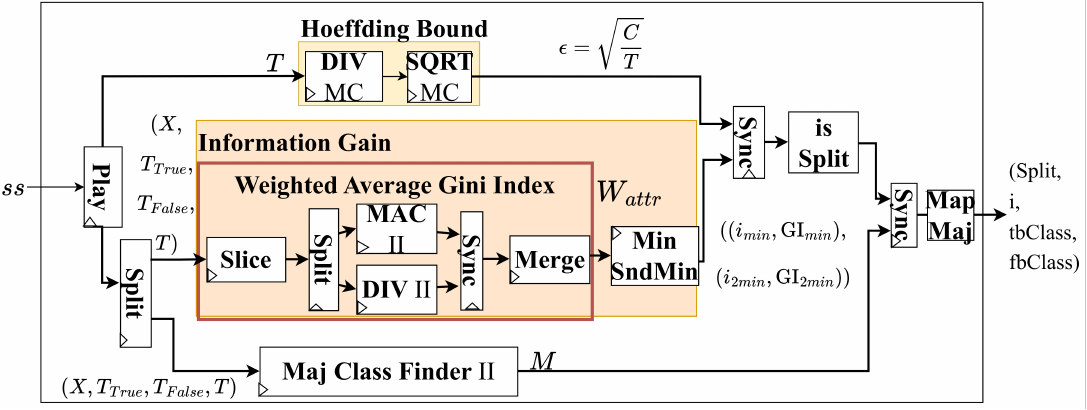}
\caption{Training dataflow architecture. Arithmetic operation configurations: multi-cycle (MC) and pipelined (II).}
    \label{fig:traindf}
\end{figure}

\begin{figure}[t]
    \centering
\includegraphics[width=\linewidth]{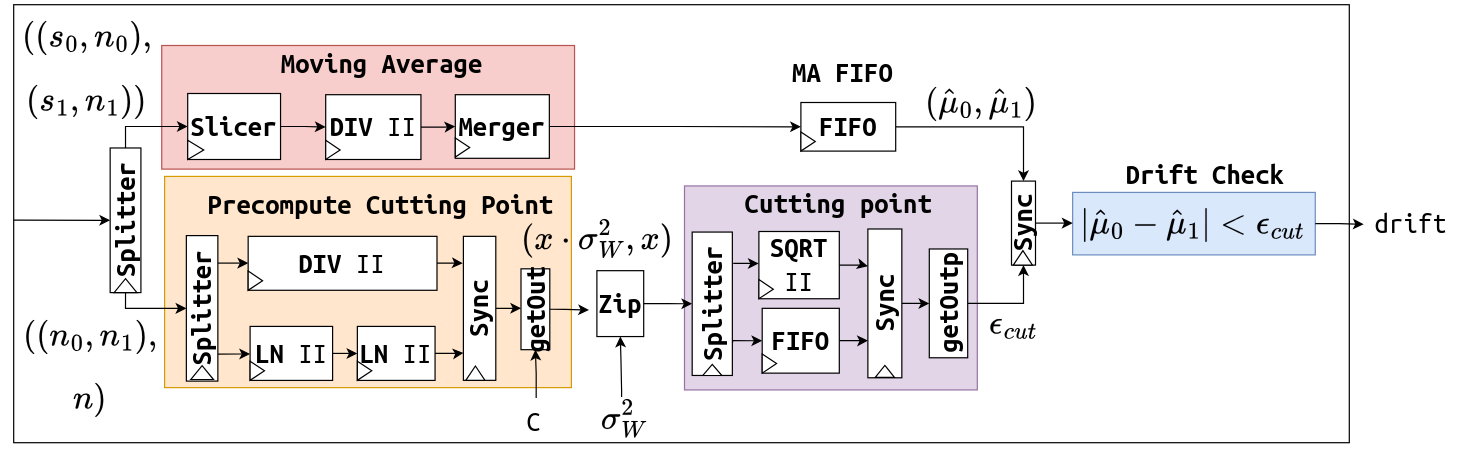}
\caption{ADWIN2 cut-check dataflow architecture.}
    \label{fig:cutcheckdf}
\end{figure}

\begin{itemize}

    \item \textbf{Training dataflow (\autoref{fig:traindf}):} consists 
    of three data paths, to which leaf statistics from the attribute subset are 
    distributed by \emph{play}. 
    The Hoeffding bound (upper \autoref{fig:traindf}) runs once per attribute subset and
    is area optimized (multi-cycle), while the information gain (middle)
    processes all attributes and is throughput optimized (pipelined).
    \gls{lut}/\gls{ff} usage is optimized by fine-tuning the branch class counter resolution.
    Furthermore, \gls{dsp} usage is optimized by: (i) reusing the middle path for 
    both the left and right branch, as the \gls{mac}'s \gls{dsp} count scales linearly with 
    the number of classes, and (ii) prescaling using bit-shifts before the \gls{mac} and division.

    \item \textbf{Drift detection dataflow (\autoref{fig:cutcheckdf}):}
    implements the ADWIN2 cut-check using an FxP-based pipelined architecture, for
    parallel processing of a set of cutting points, to sense concept drift. Both paths 
    consume the same number of inputs and thus are configured for performance.
\end{itemize}

The result is an adjustable dataflow architecture with a 
flexible resolution throughout the
data path, reducing resource consumption, 
while maintaining precision.
In addition, the modular structure makes it straightforward
to swap arithmetic algorithms (e.g. optimized division operation) 
or to derive improved task schedules through dataflow analysis,
intuitively improving performance.

\section{Implementation}\label{sec:implementation}

To validate OpenWoud's vendor-independence, scalability and performance,
we mapped the \gls{arf} onto two \glspl{fpga} from different vendors 
(Xilinx/AMD and Microsemi)
and compared it against three baseline models: 
\emph{River} (CPU) \cite{gomes_adaptive_2017}, \emph{GSARF} 
(GPU) \cite{wu_gpu-based_2020} and 
\emph{Vitis} (FPGA) \cite{ridder_accelerated_2023}.
This section goes over the hardware targets and hyperparameter alignment
to ensure a reproducible fair comparison.

\subsection{Hardware targets} 
OpenWoud is synthesized for the low-end Xilinx/AMD Kria KV260 \gls{soc} (Vivado 2022.2), 
used for on-target inference 
and Microsemi PolarFire \gls{soc} (Libero 2022.1) for cross-vendor validation.
We use River, a Python framework targeting CPUs, as baseline for accuracy measurements.
Performance metrics are omitted for River, since CPU execution
is 2 orders of magnitude slower than hardware accelerators 
such as GSARF (138x speed-up).
GSARF is implemented on a NVIDIA GeForce GTX 1650 GPU,
while Vitis (FPGA baseline) is evaluated in Vitis-HLS simulation,
targeting the high-end Alveo U55 with \gls{hbm} main memory.
Both Xilinx-\glspl{soc} have four \gls{hpb} busses between
the \gls{fpga} fabric and main memory, 
bypassing the central management unit, for high-throughput burst transfers.
However, \gls{hbm} allows for fully parallel read/write behavior, 
while \gls{ddr}-4 internally queues requests from different channels.
Both the CPU baseline and hardware accelerators use the binary 
attribute optimization, as the latter only accept binary attributes.

%
%
\subsection{Hyperparameters} 
The accelerator's hyperparameters are fine-tuned/aligned 
for each dataset (\autoref{tab:configs}). 
In this work, the considered models are restricted to $T=64$ 
trees with a depth of $D=9$,
as the smallest target (Kria) has limited \glspl{lut} and \gls{bram} depth respectively. 
Please note that the design is also scalable for larger FPGAs.
While no systematic study has established the optimal tree count
for an ARF, Oshiro et al \cite{oshiro_how_2012} show that an optimal 
RF contains between 64 and 128 trees, therefore we adopt the lower bound.
Drift detector hyperparameters: $\delta_W$ and $\delta_E$ are concept drift warning/error thresholds,
cut-check is applied periodically via the \gls{gp} to improve throughput,
$W$ is the minimum sliding-window size required to apply the cutting point algorithm, 
$M$ indicates the minimum buckets of size $2^i$ to enable compression.
Training hyperparameters: $\delta$ is hoeffding bound's confidence threshold,
tie threshold $\tau$ ensures a split after an upper bound of cycles,
\gls{gp} enables training, attribute subset size $SS$,
Poisson distribution $\lambda$.

GSARF has an additional \emph{candidate tree} type.
For fair comparison the foreground trees need to match up, 
thus GSARF has 96 trees in total,
giving it an edge over other models during concept drift recovery.
OpenWoud avoids information gain approximation (unlike GSARF/Vitis), enabling 
higher precision.

\begin{table}[t]
\scriptsize\centering
\setlength{\tabcolsep}{2pt}
\begin{tabularx}{.5\textwidth}{l l l l l l l l l l l l l l l l l l}
\toprule
\multicolumn{5}{c}{General} & \multicolumn{5}{c}{Drift Detector} & \multicolumn{5}{c}{Training-Engine} \\
\cmidrule(lr){1-5} \cmidrule(lr){6-10} \cmidrule(lr){11-15}
Dataset & ARF & Target & T & D & $\delta_W$ & $\delta_E$ & GP & W & M & $\delta$ & $\tau$ & GP & SS & $\lambda$ \\
\midrule
\textbf{AGR} & River  & CPU  & 64 & 9 & 0.001 & 0.00001 & 32 & 1000 & 5 & 0.05 & 0.05 & 16 & 10 & 1 \\
                 & GSARF  & GPU  & 96 & 9 & 0.001 & 0.00001 & 32 & 10   & 5 & 0.002 & --   & 16 & 10 & 1 \\
                 & Vitis  & FPGA & 64 & 9 & 0.001 & 0.00001 & 32 & 5    & 5 & 0.05  & 0.05 & 16 & 10 & 1 \\
                 & OpenWoud  & FPGA & 64 & 9 & 0.001 & 0.00001 & 32 & 1000 & 5 & 0.05  & 0.05 & 16 & 10 & 1 \\
\midrule
\textbf{LED}     & River  & CPU  & 64 & 9 & 0.001 & 0.00001 & 32 & 1000 & 5 & 0.05  & 0.05 & 16 & 5  & 1 \\
                 & GSARF  & GPU  & 96 & 9 & 0.001 & 0.00001 & 32 & 10   & 5 & 0.002 & --   & 16 & 5  & 1 \\
                 & Vitis  & FPGA & 64 & 9 & 0.001 & 0.00001 & 32 & 5    & 5 & 0.05  & 0.05 & 16 & 5  & 1 \\
                 & OpenWoud  & FPGA & 64 & 9 & 0.001 & 0.00001 & 32 & 1000 & 5 & 0.05  & 0.05 & 16 & 5  & 1 \\
\midrule
\textbf{COV}     & River  & CPU  & 64 & 9 & 0.001 & 0.00001 & 32 & 1000 & 5 & 0.05  & 0.05 & 16 & 7  & 1 \\
                 & GSARF  & GPU  & 96 & 9 & 0.001 & 0.00001 & 32 & 10   & 5 & 0.002 & --   & 16 & 7  & 1 \\
                 & Vitis  & FPGA & 64 & 9 & 0.001 & 0.00001 & 32 & 5    & 5 & 0.05  & 0.05 & 16 & 7  & 1 \\
                 & OpenWoud  & FPGA & 64 & 9 & 0.001 & 0.00001 & 32 & 1000 & 5 & 0.05  & 0.05 & 16 & 7  & 1 \\
\bottomrule
\end{tabularx}
\caption{ARF hyperparameters per dataset and model.}
\label{tab:configs}
\end{table}

\section{Evaluation}\label{sec:evaluation}

This section presents a comprehensive quantitative evaluation of OpenWoud,
in the form of a benchmark test.

\subsection{Experimental Setup}

To evaluate OpenWoud, an infrastructure is developed to apply
various evolving data streams to the ARF.

%
%
\textbf{Data movers}
A \gls{dma} controller moves
samples and votes in bulk between the \gls{arf} and main memory, through
\gls{axi} \gls{hpb} interfaces, minimizing
software overhead and additional design complexity 
(e.g. caused by an external stream generator).
Its read/write channels are multiplexed using a smart-connect on HP1 and HP2,
while the training-engine uses HP0 and HP3 to update leaf statistics.
The hard-processor of the \gls{soc} is used for: 
configuring \gls{dma} transfers, sample/vote processing
and polling completion flags.
All \gls{fpga} experiments run with a system clock of 100MHz.

%
%

%
%

\begin{table}[t]
    \centering\footnotesize
    \begin{tabularx}{.45\textwidth}{r r c c c}
        \hline
        \textbf{Dataset} & \textbf{Type} & \textbf{Samples} & \textbf{Attributes} & \textbf{Classes} \\
        \hline
        AGRAWAL & Synthetic & 1,000,000 & 82 & 2 \\
        LED     & Synthetic & 1,000,000 & 24 & 10 \\
        COVTYPE & Real      & 581,012   & 44 & 7 \\
        KDD99   & Real      & 494,020   & 24 & 23 \\
        \hline
    \end{tabularx}
    \caption{Synthetic and real datasets.}
    \label{tab:datasets}
\end{table}

\textbf{Datasets}
Benchmark tests are performed using two synthetic datasets
called AGRAWAL by Agrawal et al \cite{agrawal_database_1993} and 
LED by Breiman et al \cite{breiman_random_2001} 
(each contains abrupt, gradual and mixed drift)
and, two real datasets: 
Forest cover type (COVTYPE) by Blackard et al 
\cite{blackard_covertype_1998}, and 
KDD Cup 1999 (KDD-99) for network intrusion detection 
by Hettich et al \cite{hettich_kdd_1999}.
These four sets allow us to study both concept drift recovery and
real-world classification performance,
in line with the FPGA baseline.
Attribute preprocessing is performed with conversion 
scripts\footnote{https://github.com/ingako/gsarf} by Wu et al.


\begin{figure*}[!htbp]
  \centering
  \includegraphics[width=\linewidth]{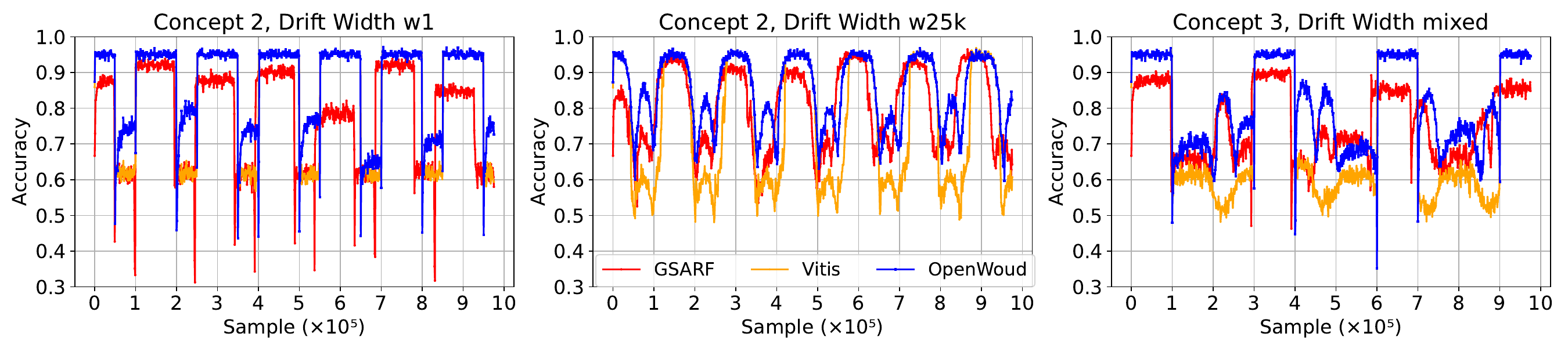}
  \caption{AGRAWAL (synthetic) drift recovery comparison 
  GSARF (GPU) \cite{wu_gpu-based_2020},
  Vitis (GPU) \cite{ridder_accelerated_2023} and 
  OpenWoud (FPGA).
  }
  \label{fig:agrawalacc}
\end{figure*}

\begin{figure*}[!htbp]
  \centering
  \includegraphics[width=\linewidth]{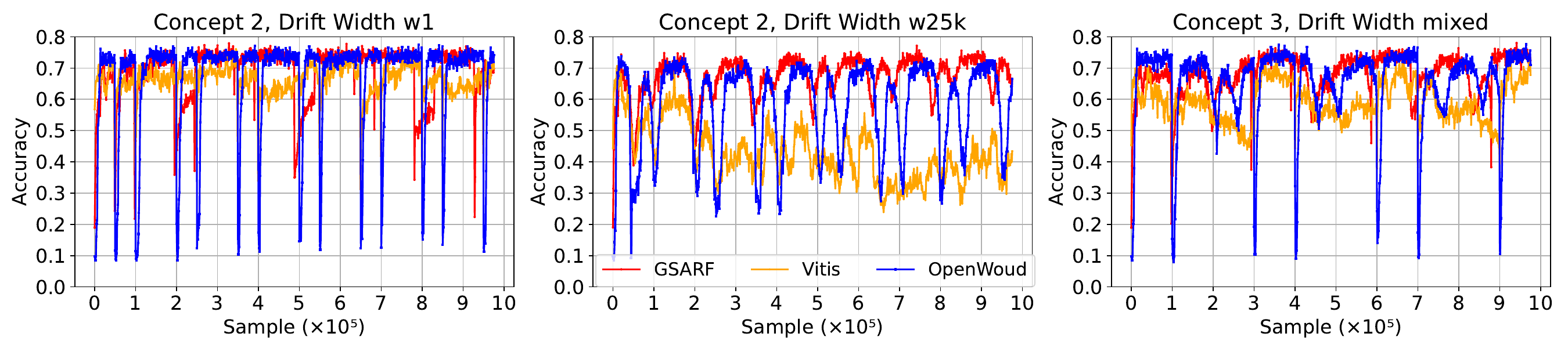}
  \caption{LED (synthetic) drift recovery comparison 
  GSARF (GPU) \cite{wu_gpu-based_2020},
  Vitis (GPU) \cite{ridder_accelerated_2023} and 
  OpenWoud (FPGA).
  }
  \label{fig:ledacc}
\end{figure*}

%
%
\subsection{Area footprint}

%
%
Compared to the FPGA baseline, OpenWoud requires:
52\% \gls{lut}, 24\% \gls{ff}, 77\% \gls{dsp} and 40\% \gls{bram} 
resources on average, while eliminating \gls{uram} usage 
(Xilinx-specific building-block).
Enabling deployment on smaller Xilinx-FPGAs, where the FPGA baseline
fails to fit (3 out of 4 datasets on Kria \autoref{tab:kria-area}) and
non-Xilinx (all datasets on PolarFire \autoref{tab:polarfire_resources}).
Crucially, this cross-vendor portability allows OpenWoud 
to exploit build-in fabric features of different families,
including radiation-tolerant platforms, widely used
in the aerospace domain, to increase fault-resilience.

A closer look at the resource breakdown shows that \glspl{lut} and
\gls{bram} are the most heavily utilized building blocks.
\glspl{lut} primarily 
in the training and drift detection components, to implement multiplexing logic
for leaf statistics and sliding-window indexing.
While tree and histogram storage obtain most of the \glspl{bram}.

In addition, OpenWoud provides improved scalability
over the FPGA baseline, when adapting 
hyperparameters such as: tree count/depth, attributes and classes,
resource usage grows predictably, fine-tuning the baseline 
is less intuitive, due to its reliance on \gls{hls} pragmas and compiler-driven, 
non-transparent scheduling.

\begin{table}[b]

\centering
\scriptsize
\begin{tabularx}{\linewidth}{l c c c c c c} 
\toprule
\textbf{SET} & \textbf{MODEL} & \makecell{\textbf{6LUT}\\(117120)} & \makecell{\textbf{DFF}\\(234240)} & \makecell{\textbf{DSP}\\(1248)} & \makecell{\textbf{BRAM}\\(144)} & \makecell{\textbf{URAM}\\(64)} \\
\midrule
\multirow{2}{*}{AGR} 
    & OpenWoud & 87046 & 24761 & 143 & 112 & 0 \\
    & Vitis & \textcolor{red}{180974} & 109229 & 191 & \textcolor{red}{404} & 16 \\
\addlinespace
\multirow{2}{*}{LED} 
    & OpenWoud & 83803 & 26709 & 155 & 112 & 0 \\
    & Vitis & \textcolor{red}{159825} & 120439 & 192 & \textcolor{red}{211} & 16 \\
\addlinespace
\multirow{2}{*}{COV} 
    & OpenWoud & 107322 & 29349 & 150 & 112 & 0 \\
    & Vitis & \textcolor{red}{189813} & 140074 & 196 & \textcolor{red}{288} & 16 \\
\addlinespace
\multirow{2}{*}{KDD} 
    & OpenWoud & \textcolor{red}{149885} & 54378 & 140 & 72 & 0 \\
    & Vitis & \textcolor{red}{231094} & \textcolor{red}{278763} & 94 & 64 & 8 \\
\bottomrule
\end{tabularx}
\caption{OpenWoud resource utilization report Kria.}
\label{tab:kria-area}
\end{table}

\begin{table}[b]
\centering
\scriptsize
\begin{tabularx}{\columnwidth}{>{\raggedleft\arraybackslash}X *{8}{>{\centering\arraybackslash}X}}
\toprule
\textbf{SET} & \textbf{MODEL} & \textbf{Fab 4LUT} & \textbf{Fab DFF} & \textbf{Intf 4LUT} & \textbf{Intf DFF} & \textbf{Math} & \textbf{uSRAM} & \textbf{LSRAM} \\
    & & \texttt{(254k)} & \texttt{(254k)} & \texttt{(254k)} & \texttt{(254k)} & \texttt{(784)} & \texttt{(2.4k)} & \texttt{(812)} \\
\midrule
AGR & T64 & 13,691 & 29,704 & 14,292 & 14,292 & 297 & 128 & 170 \\
LED     & T64 & 123,572 & 31,424 & 14,616 & 14,616 & 297 & 128 & 179 \\
COV & T64 & 151,092 & 35,622 & 14,508 & 14,588 & 297 & 128 & 176 \\
KDD   & T32 & 224,700 & 50,886 & 11,244 & 11,244 & 169 & 64 & 192 \\
\bottomrule
\end{tabularx}
\caption{OpenWoud resource utilization report PolarFire.}
\label{tab:polarfire_resources}
\end{table}

%
%
\subsection{Accuracy}

%
%
OpenWoud maintains or improves precision across all datasets (\autoref{tab:benchmark}),
average accuracy differences compared to 
River (CPU) / GSARF (GPU) / Vitis (FPGA) respectively:

\begin{itemize}
    \item AGRAWAL (synthetic): -0.97\%/+6.3\%/+8.45\%
    \item LED (synthetic): -7.11\%/-4.56\%/+6.55\%
    \item COVTYPE (real): -2.93\%/+12.89\%/+7.31\%
\end{itemize}

The improvement stems from fast precise FxP dataflows,
avoiding information gain approximations (unlike GSARF/Vitis).
As a result, OpenWoud maintains baseline CPU accuracy,
while still benefitting from hardware acceleration.

%
%
We evaluate the concept drift recovery behavior of the \gls{arf}
on synthetic AGRAWAL and LED datasets,
for abrupt (w1), gradual (w25k) and mixed drift 
(\autoref{fig:agrawalacc} and \autoref{fig:ledacc})

\begin{table*}[t]
\scriptsize
\setlength{\tabcolsep}{2pt}
\begin{tabularx}{\textwidth}{c c l
    *{4}{>{\centering\arraybackslash}X}
    *{3}{>{\centering\arraybackslash}X}
    *{2}{>{\centering\arraybackslash}X}
}
\toprule
Dataset & \#Concepts & Drift Width
& \multicolumn{4}{c}{Accuracy [\%]} 
& \multicolumn{3}{c}{Inference Time} 
& \multicolumn{2}{c}{OpenWoud Speed Up Factor} \\
\cmidrule(lr){4-7} \cmidrule(lr){8-10} \cmidrule(lr){11-12}
& & & River\newline(CPU) & GSARF (GPU) & Vitis\newline(FPGA) & OpenWoud (FPGA)
& GSARF & Vitis & OpenWoud
& vs GSARF & vs Vitis \\
\midrule
\multirow{6}{*}{AGRAWAL}
& 2 & 1     & 86.45 $\pm$ 0.26 & 79.04 $\pm$ 1.84 & 83.21 $\pm$ 0.06 & 86.69 $\pm$ 0.46 & 16.11s & 9.1s  & 25.9s  & 0.6 & 0.4 \\
& 2 & 25k   & 83.87 $\pm$ 0.21 & 80.17 $\pm$ 1.08 & 75.67 $\pm$ 0.24 & 85.24 $\pm$ 0.46 & 22.66s & 9.33s & 24.26s & 0.9 & 0.4 \\
& 2 & 50k   & 81.28 $\pm$ 0.14 & 77.62 $\pm$ 0.74 & 68.93 $\pm$ 0.36 & 75.23 $\pm$ 0.53 & 22.99s & 9.33s & 20.34s & 1.1 & 0.5 \\
& 3 & 1     & 87.31 $\pm$ 0.24 & 74.4 $\pm$ 3.07  & 77.62 $\pm$ 0.34 & 88.99 $\pm$ 0.22 & 16.43s & 9.12s & 25.92s & 0.6 & 0.4 \\
& 3 & 25k   & 81.87 $\pm$ 0.21 & 73.57 $\pm$ 1.76 & 69.56 $\pm$ 0.27 & 80.55 $\pm$ 0.37 & 22.78s & 9.34s & 24.76s & 0.9 & 0.4 \\
& 3 & mixed & 83.69 $\pm$ 0.21 & 76.03 $\pm$ 1.22 & 72.98 $\pm$ 0.1  & 81.95 $\pm$ 0.25 & 19.76s & 9.24s & 25.63s & 0.8 & 0.4 \\
\midrule
\multirow{6}{*}{LED}
& 2 & 1     & 72.87 $\pm$ 0.04 & 69.6 $\pm$ 1.26  & 65.51 $\pm$ 1.63 & 67.13 $\pm$ 0.25 & 42.99s & 9.87s & 42.78s & 1.0 & 0.2 \\
& 2 & 25k   & 67.82 $\pm$ 0.05 & 67.32 $\pm$ 0.4  & 47.8 $\pm$ 2.89  & 58.23 $\pm$ 0.41 & 47.51s & 9.52s & 42.07s & 1.1 & 0.2 \\
& 2 & 50k   & 64.06 $\pm$ 0.15 & 64.15 $\pm$ 0.65 & 42.81 $\pm$ 3.05 & 57.68 $\pm$ 0.42 & 48.06s & 9.37s & 31.49s & 1.5 & 0.3 \\
& 3 & 1     & 73.16 $\pm$ 0.06 & 67.49 $\pm$ 0.77 & 65.38 $\pm$ 1.54 & 66.62 $\pm$ 0.22 & 41.34s & 9.12s & 42.95s & 1.0 & 0.2 \\
& 3 & 25k   & 69.54 $\pm$ 0.05 & 66.51 $\pm$ 0.37 & 55.9 $\pm$ 1.65  & 61.88 $\pm$ 0.21 & 49.37s & 9.45s & 42.15s & 1.2 & 0.2 \\
& 3 & mixed & 71.55 $\pm$ 0.06 & 68.64 $\pm$ 0.53 & 59.65 $\pm$ 1.16 & 65.45 $\pm$ 0.17 & 47.34s & 9.78s & 42.3s  & 1.1 & 0.2 \\
\midrule
COVTYPE & - & - & 72.8 & 56.98 & 62.56 & 69.87 & 12.71s & 6.78s & 25.24s & 0.5 & 0.3 \\
\bottomrule
\end{tabularx}
\caption{OpenWoud performance and precision benchmarks.}
\label{tab:benchmark}
\end{table*}

%
%
On AGRAWAL, OpenWoud consistently achieves faster and more stable 
recovery than the GPU and FPGA baseline.
When a new concept occurs, the accuracy drops sharply across all models,
OpenWoud goes back to pre-drift levels quicker.

On LED, the GPU shows an advantage on recurring concepts, where
its candidate tree mechanism yields quicker recovery.
In contrast, OpenWoud experiences slightly deeper accuracy dips,
but still recovers steadily.
While the FPGA baseline collapses completely on gradual concept drift.

Overall, these results demonstrate that OpenWoud provides 
robust recovery of concept drift, while 
sustaining its vendor-independence and sample-by-sample processing.

\subsection{Inference-time}

We measure OpenWoud's end-to-end inference-time (\autoref{tab:benchmark}) 
from the \gls{dma} bulk transfer start signal, 
till the \gls{dma}-finished interrupt on the hard-processor.
OpenWoud achieves competitive performance compared to the 
GPU (0.5 - 1.5$\times$ speed-up) 
and FPGA baseline (0.2 - 0.5$\times$).

Importantly, the reported FPGA baseline is not measured on hardware,
but derived analytically \cite{ridder_accelerated_2023},
using the active average number of foreground trees, obtained through simulation.
In addition, the calculation assumes that main memory does not queue requests,
this is true for \gls{hbm}, however \gls{ddr}-4 (available on Kria) cannot deliver 
truly parallel read/write behavior across multiple interfaces.
Furthermore, the GPU requires batch-based sample insertion, while
the FPGA models sustain true sample-by-sample inference,
required for low latency
on evolving data streams for real-time applications.
\section{Conclusion}\label{sec:conclusion}
%
%
We presented OpenWoud, a vendor-unlocked FPGA accelerator
for the \gls{arf}. 
By introducing a flexible \gls{fxp} arithmetic library,
and fully transparent, manually scheduled dataflow architectures,
OpenWoud overcomes the scalability limitations of the FPGA baseline.

Our quantitative evaluation (benchmark test) shows OpenWoud maintains or increases 
accuracy (up to 12.89\%) over existing hardware accelerators.
It sustains competitive throughput, delivers
true sample-by-sample processing and robust drift recovery.
Importantly, OpenWoud scales predictably when adjusting hyperparameters,
avoiding non-intuitive pragma tuning and opaque task scheduling
of \gls{hls}-based baselines.
These results demonstrate that OpenWoud provides a practical 
and portable framework, for real-time online \gls{ml}.

By mapping successfully onto both Xilinx and MicroSemi FPGAs, 
OpenWoud validates its cross-vendor portability. 
This capability allows it to exploit vendor-specific fabric features,
such as radiation-tolerant flash-based solutions from MicroSemi,
making the framework especially relevant for aerospace research,
where fault-resilience is critical.

Looking ahead, several research directions remain open.
First, investigating the \gls{arf}'s fault-resilience 
for aerospace applications, through a 
radiation-hardening campaign on flash-based platforms.
Second, exploring the \gls{arf}'s power efficiency,
which is particularly promising on \gls{fpga}
accelerators due to fine-grained clock gating.
Third, deriving Pareto-optimal resolutions 
for time/area trade-offs in dataflows,
and extending this with dataflow analysis, 
to optimize the task schedule. 

\IEEEtriggeratref{14}
\bibliographystyle{IEEEtran}
\bibliography{references.bib}
\end{document}